\documentclass[runningheads]{llncs}
\usepackage{graphicx}
\usepackage{amsmath}
\usepackage{booktabs}
\usepackage{makecell}
\begin{document}

%
% \title{GAN-based Perceptual Image Compression with DISTS Optimization}
\title{GAN-based Image Compression with Improved RDO Process}
\author{Fanxin Xia\inst{1}
\and
Jian Jin\inst{2}
\and
Lili Meng\inst{1}
\and
Feng Ding\inst{1}
\and
Huaxiang Zhang\inst{1}}
\authorrunning{X. Fanxin et al.}
\institute{Information Science and Engineering, Shandong Normal University, Jinan, 250014, China.\\
\email{2021020980@stu.sdnu.edu.cn}
%\email{fengding96@gmail.com}\\
%\email{\{mengll\_83,huaxzhang\}@hotmail.com}
\and
Computer Science and Engineering, Nanyang Technological University, 639798, Singapore\\
% \email{ jian.jin@ntu.edu.sg}
}
\maketitle              
% typeset the header of the contribution
%
\begin{abstract}
% GAN-based image compression schemes have shown remarkable progress lately due to their high perceptual quality at low bit rates. However, there are three challenges, including sufficient texture details, training stability, and reasonable perceptual optimization metric. In this paper, we present a novel GAN-based image compression approach that addresses these challenges above. We combine MS-SSIM and DISTS metrics in the rate-distortion optimization function to achieve a balanced emphasis on GAN training and perceptual reconstruction capability. We also absorb the discretized gaussian-laplacian-logistic mixture model to improve the accuracy of entropy estimation in entropy modeling. In addition, we conducted a Mean Opinion Score (MOS) experiment using a more accurate subjective quality assessment method, as current quality metrics do not fully represent the actual perceptual results of humans. Experimental results demonstrate that our method outperforms existing GAN-based methods and compression standards in perceptual quality. 

GAN-based image compression schemes have shown remarkable progress lately due to their high perceptual quality at low bit rates. However, there are two main issues, including 1) the reconstructed image perceptual degeneration in color, texture, and structure as well as 2) the inaccurate entropy model. In this paper, we present a novel GAN-based image compression approach with improved rate-distortion optimization (RDO) process. To achieve this, we utilize the DISTS and MS-SSIM metrics to measure perceptual degeneration in color, texture, and structure. Besides, we absorb the discretized gaussian-laplacian-logistic mixture model (GLLMM) for entropy modeling to improve the accuracy in estimating the probability distributions of the latent representation. During the evaluation process, instead of evaluating the perceptual quality of the reconstructed image via IQA metrics, we directly conduct the Mean Opinion Score (MOS) experiment among different codecs, which fully reflects the actual perceptual results of humans. Experimental results demonstrate that the proposed method outperforms the existing GAN-based methods and the state-of-the-art hybrid codec (\emph{i.e.}, VVC).

\keywords{Perceptual image compression  \and GAN-based method.}
\end{abstract}

\section{INTRODUCTION}
Image compression is an essential technique for effective image representation, storage, transmission, and so on. With the great successes of deep learning in computer vision and signal processing, many learning based image compression schemes have been developed recently. Toderici \emph{et al}. \cite{toderici2015variable} proposed a learnable image compression method based on the recurrent neural network (RNN) to disentangle thumbnail image compression. After that, Ballé \emph{et al}. \cite{balle2016end} developed the first end-to-end image compression framework, which was mainly made up of convolutional autoencoder and mixed with Generalized Divisive Normalization (GDN) \cite{J2015Density}. Besides, Ballé \emph{et al}. \cite{balle2018variational} further developed a hyperprior model to extract side information for estimating the probability of the latent representation. Although image compression schemes like \cite{balle2016end1,theis2017lossy,minnen2018joint} have made significant advancements in PSNR and MS-SSIM, the biggest challenge still focuses on achieving high perceptual quality image reconstruction, especially for image compression under the low bit rates. %As shown in Fig. 1, the value of existing evaluation indicators can not completely correspond to the quality of human perceptions.

\begin{figure}[t]
\centering
\includegraphics[width=12cm]{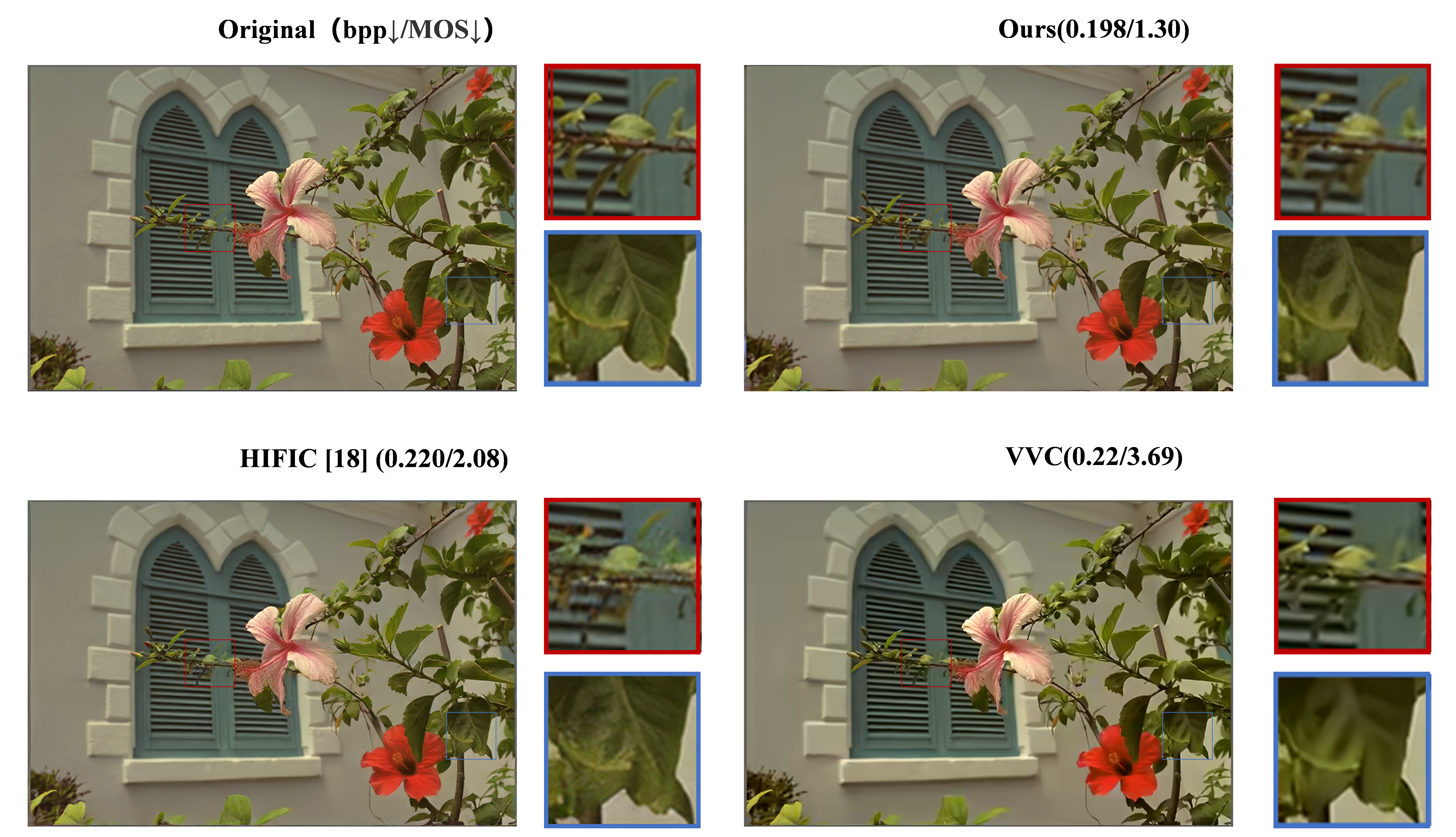}
\caption{Visual comparison among the original image, compressed images with our method, HIFIC \cite{mentzer2020high} (state-of-the-art GAN-based image compression method), and VVC \cite{9503377} (state-of-the-art hybrid image codec). The bpp and MOS values of each compressed image are exhibited as well.}
\label{MOS}
\end{figure} 

To overcome this problem, some works \cite{cheng2021perceptual,agustsson2019generative,iwai2021fidelity,mentzer2020high} tried to utilize the Generative Adversarial Network (GAN) \cite{creswell2018generative} to replace the decoder for reconstructing high perceptual quality images. For example, Agustsson \emph{et al}. \cite{agustsson2019generative} proposed a method that used GAN to achieve image compression at extremely low bit rates and demonstrated satisfactory results. 
However, this method did not adequately consider the perceptual quality degeneration in the reconstructed image. Fabian \emph{et al}. \cite{mentzer2020high} proposed a perceptual image compression model that incorporated LIPIS \cite{2018The} together with MSE into the RDO loss function. However, LIPIS overlooked the essential factor of color, which caused the perceptual quality degradation of the reconstructed image in color and texture. Besides, MSE is commonly used for maintaining the fidelity of signal, which overlooked the structure information of reconstructed image. It has been demonstrated in \cite{wang2004image} that structure information is one of the most sensitivity factors for the perceptual quality of humans vision. Hence, more suitable perceptual metrics are expected to be utilized for perceiving the distortion in the reconstructed image. %Based on this, our proposed method utilizes DISTS \cite{9298952} and MS-SSIM metrics to measure perceptual degeneration in many aspects, such as texture, color, detail, and structure, which will improve the accuracy and reliability of image quality assessment, and the reconstruction can better match with human perceptual judgments.
% Based on this, our proposed method utilizes DISTS \cite{9298952} as a perceptual loss that considers texture, color, and human perception. And we replace MSE with MS-SSIM, which is more in line with human perception, to achieve better results while maintaining stable training.

In addition, the entropy model, used for estimate bit rate cost in the RDO, is also an important part of GAN-based image compression. However, the accuracy of entropy model in most of existing GAN-based image compression works was low, which also affect the performance of image compression. For instance, the discretized Gaussian mixture model (GMM) and zero-mean Gaussian distribution model were used in \cite{cheng2021perceptual,iwai2021fidelity} and \cite{mentzer2020high} for entropy modeling, which were difficult to accurately estimate the probability distribution of each image. Therefore, a more accurate probability model is expected to be used for entropy modeling.

%our method utilizes the discretized Gaussian-Laplacian-Logistic mixture model (GLLMM) \cite{fu2021learned} to achieve more accurate probability estimation and better results.

In this paper, our main contributions are as follows:
\begin{itemize}
  \item We integrate DISTS and MS-SSIM into the RDO loss function, which provides a comprehensive assessment of distortion in the reconstructed image in terms of color, texture, and structure.
  % We choose DISTS as the perceptual loss, which provides a comprehensive assessment of different aspects of the image. Additionally, we utilized MS-SSIM to stabilize the training, and it is better suited to optimizing the GAN-based image compression method.

  \item We absorb a discretized GLLMM probability model for entropy modeling and achieve higher accuracy in estimating the probability distributions of the latent representation, reducing the reconstruction error attributed to imprecise entropy estimation. 
  
  \item Subjective tests demonstrate the proposed method has better perceptual quality and lower BPP cost compared with the existing state-of-the-art GAN-based image codec and the hybrid one (i.e., VVC), as shown in Fig. \ref{MOS}. 
\end{itemize}

\section{GAN-based Image Compression Method Review}
\begin{figure}[htb]
\centering
\includegraphics[width=7cm]{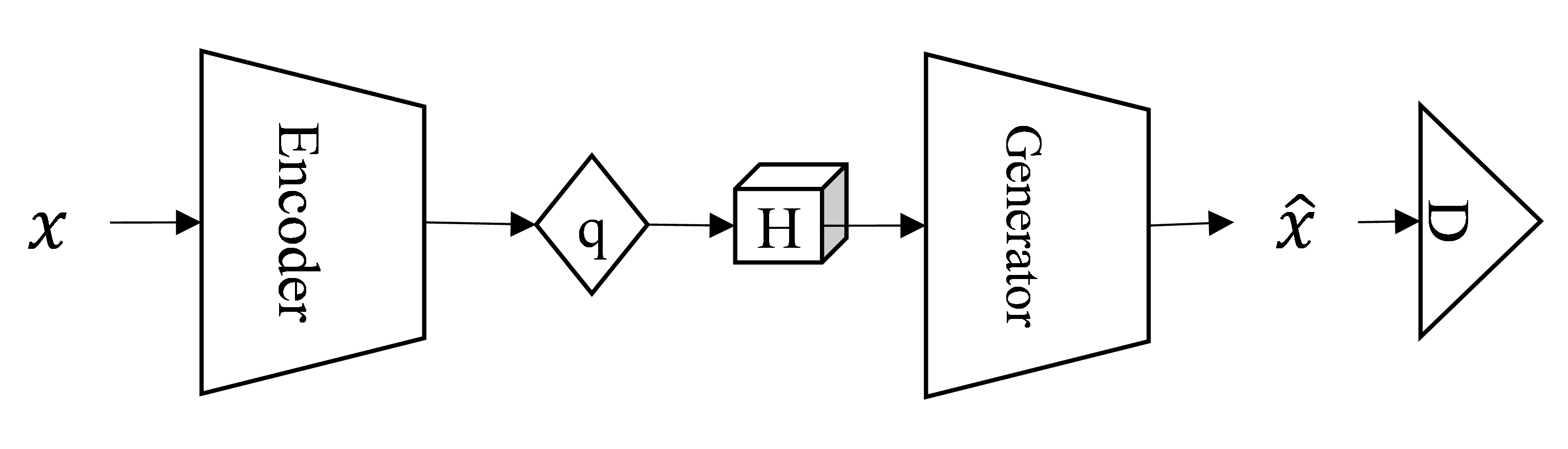}
\caption{The framework of the GAN-based image compression. The  image $x$ to be compressed is fed into the encoder at first. Then, it is represented with latent $y$. Afterward, $y$ is quantized and entropy coded to $\omega=H(q(y))$. Then, it is decoded by the generator to reconstruct the image $\hat{x}$. The input image $x$ and its corresponding reconstructed one $\hat{x}$ are alternately fed into a discriminator to improve its discriminative capability. Meanwhile, to fool the discriminator, the generator is further improved to produce higher-quality images.
}
\label{GAN}
\end{figure}

\subsection{GAN}

GAN typically consists of two components, one is the generator and the other is the discriminator. The objective function of the GAN model is as follows.

\begin{equation}
\mathop{\min}_{G}\mathop{\max}_{D} V(D,G)={E}[log(1-D(G(z)))]+E[log(D(x))],
\end{equation}
where \emph{x} and \emph{z} denote the real image and the latent representation, respectively. ${E}[log(1-D(G(z)))]$ is the objective function for the generator, which is used to produce examples that has genuine and authentic content to the original ones. $E[log(D(x))]$ is the objective function for the discriminator, which is used to train a powerful discriminator to distinguish the generated examples from the original ones. By combining these two items, the generator is able to generate examples that are close enough to the original ones. Hence, GAN technology has been applied to a range of tasks in recent years, including  image-inpainting \cite{yu2019free}, image deblurring \cite{kupyn2018deblurgan}, image compression \cite{iwai2021fidelity}, and so on.

\subsection{GAN-Based Image Compression}
\label{GIC}

Fig. \ref{GAN} illustrates the framework of a typical GAN-based image compression method, which is mainly made up of an encoder, a generator, a discriminator, and an entropy model. The goal of integrating GAN into the learning-based image compression method was to utilize its powerful generator for high-quality image reconstruction. Such as \cite{9150785}, the authors incorporated GAN into the original structure as a reconstruction enhancement method to improve the perceptual reconstruction capability. 
Furthermore, GAN's strong generation capability has made it a common choice for image compression methods \cite{kim2020towards,akutsu2020ultra,9093387} targeting extremely low bitrates to overcome the problem of insufficient details in low bitrates.

The optimization process still follows the RDO, but the incorporated GAN needs additional adversarial losses to optimize the generator and discriminator, respectively. The loss function of the GAN-based image codec is formulated as follows.

\begin{equation} \label{GAN-loss}
\mathop{\min}_{E,G}\mathop{\max}_{D} V(D,G)={E}[f(D(x))]+E[g(D(G(z)))]+\lambda{E}(d(x, G(z))+\beta H(\hat{y}),
\end{equation}
which four items. The first two items refer to the GAN losses for discriminator and generator, respectively. ${E}(d(x, G(z))$ and $H(\hat{y})$ are the rate-distortion loss based on Shannon theory \cite{cover1999elements}. To make the training of GAN stable, the instability of GAN network training, the $d(x, G(z))$ term often needs MSE to avoid mode collapse.

\begin{figure}[t]

\centering
\includegraphics[width=12cm]{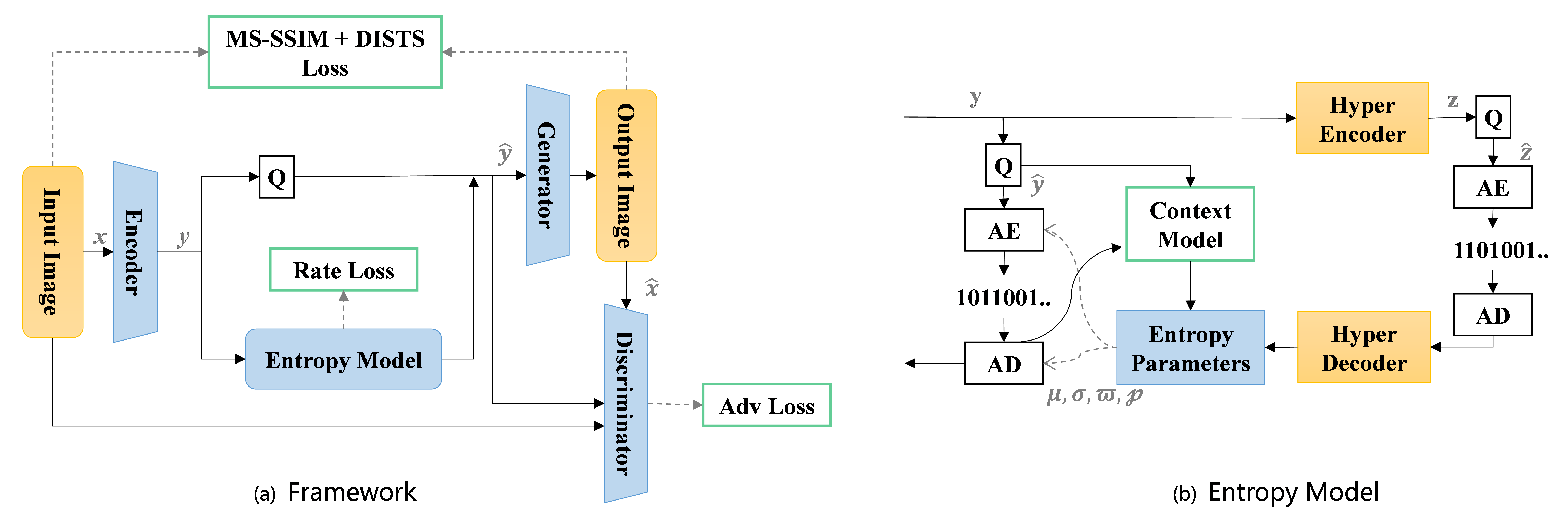}
\caption{The framework of the proposed method. First, image $x$ is fed into the encoder to achieve latent representation $y$. Then, $y$ is quantized into $\hat{y}$ through rounding. Meanwhile, $y$ is fed into the entropy model to estimate the probability distribution of each entry. Next, $\hat{y}$ is fed into the generator for image reconstruction. Finally, $\hat{y}$ is fed into the discriminator along with the original image $x$ and the reconstructed image $\hat{x}$. $Q$, $AE$, and $AD$ are a quantizer, an arithmetic encoder and an arithmetic decoder, respectively.
}
%Q denotes the rounding operation. The discrete elements $\hat{y}$ for image reconstruction input into the discriminator are quantized through rounding operation. Another branch means that the latent representation $y$ also undergoes the normal entropy coding for completing the RDO process. To ensure high generation quality, the discrete elements $\hat{y}$, along with the original and generated images, are fed into the discriminator to help the training. The green color highlights the loss function that we use.}
\label{struct}
\end{figure}
\section{METHOD}

% The network architecture we used is shown in Fig.3, which follows the state-of-the-art GAN-based image compression method \cite{mentzer2020high}. It contains an encoder, a generator, a discriminator, and a hyperprior coding model. Our approach differs from \cite{mentzer2020high} in three ways. Firstly, we incorporate DISTS and MS-SSIM into the loss function to enhance the perceptual quality. Secondly, we employ the discretized gaussian-laplacian-logistic mixture model as the entropy estimation model. Thirdly, we replace some of the activation functions with Leaky-Relu to increase the scope of data processing 

% The specific structure is as follows:

%\subsection{Encoder and Generator}
\subsection{Network Architecture}
The framework of our proposed method is shown in Fig. 3 (a), which is a variant of the state-of-the-art GAN-based image compression method \cite{mentzer2020high}. It contains an encoder, a generator, a discriminator, and a hyperprior coding model. The detailed structure is as follows.
\\
\textbf{Encoder and Generator.}
The encoder is made up of six convolution layers, and the generator contains five residual blocks and the de-convolution layers. Meanwhile, the ChannelNorm is carried out after each layer of encoder and generator to reduce the darkening artifacts. Instead of directly using the activation function ``Relu'' in the encoder and the generator in \cite{mentzer2020high}, ``Leaky-Relu'' is used to expand the scope of features processing, makes data processing more accurate, and enables the network to accelerate convergence. 
\\
\textbf{Discriminator.}
The purpose of the discriminator is to help the generator produce more realistic images. As shown in Fig. 3 (a), we feed $\hat{y}$ together with the original image $x$ and its corresponding reconstructed one $\hat{x}$ into the discriminator, which is made up of five convolution layers. %This approach enhances the discriminator's ability to accurately determine the authenticity of the images, resulting in improved quality of the generated images. 

% Table generated by Excel2LaTeX from sheet 'Sheet2'
\begin{table}[htbp]
\renewcommand\arraystretch{1}
  \centering
  \caption{The network struct we used, where all the layers use Leaky-Relu as activation function except for the last layer with no activation.}\label{tab1}
    % \scalebox{0.51}
    {
    \begin{tabular}{p{3.5cm}|p{5cm}|p{3.5cm}}
    \toprule
    {Encoder} & {Generator} & {Discriminator} \\
    \hline
    {Conv 7×7×60 Norm} & {Norm Conv 960 Norm} & {Conv 12 - NN 16} \\
    \hline
    {Conv 120 s2 Norm} & {ResBlock(×5): Conv 960 Norm} & {Conv 4×4×64 s2} \\
    \hline
    {Conv 240 s2 Norm} & {Conv 480 s2 Norm} & {Conv 4×4×128 s2} \\
    \hline
    {Conv 480 s2 Norm} & {Conv 240 s2 Norm} & {Conv 4×4×256 s2} \\
    \hline
    {Conv 960 s2 Norm} & {Conv 120 s2 Norm} & {Conv 4×4×512 } \\
    \hline
    {Conv 220} & {Conv 60 s2 Norm} & {Conv 1×1×1 Sigmoid} \\
    \hline
    {} & {Conv 7×7×3} & {} \\
    \bottomrule
    \end{tabular}}

\end{table}

% we use \^{y} as a condition for the discriminator, not only the original and reconstructed images, as is typical for conditional GAN. Specifically, we concatenate \^{y} with the original image and the reconstructed image, which has been upsampled sixteen times, and then fed into the discriminator. We employ Leaky-Relu as the activation function for each layer and apply Space-Norm within this section.

\subsection{Entropy Model}

The entropy model plays a critical role in the compression process, as its accuracy directly affects the probability distribution of each compressed entry and further affects the quality of the reconstructed images. A recent study in \cite{fu2021learned} has demonstrated that a single distribution probability model is not sufficient for representing the complex distribution of large images. In order to improve the accuracy, we replace the zero-mean Gaussian distribution model with the discretized GLLMM probability model in \cite{fu2021learned} for entropy modeling. The details of the entropy model are exhibited in Fig. 3 (b).

Here, the probability of each entry in core latent representation $\hat{y}$ and the side information $\hat{z}$ are to be estimated. To achieve this, we use a non-parametric and fully factorized density model in \cite{balle2018variational} to estimate the probability of $\hat{z}$, which is formulated as,

\begin{equation}
p_{\hat{z}|\psi}(\hat{z}|\psi)=\prod_{i}(p_{z_i}|\psi(\psi)\ast U(-\frac{1}{2},\frac{1}{2}))(\widehat{z_i}),
\end{equation}
where $z_i$ is the $i$-th entry of the side information $z$, $i$-th represents the location index in the feature tensor. $\psi$ is the parameters of each univariate distribution $p_{z_i}|\psi$.

For quantized latent representation $\hat{y}$, the discretized GLLMM probability model with different mean and variance is used for estimation. The entropy model is thus formulated as
\\
\begin{equation}
\begin{split}
P_{\hat{y}|\hat{z}}(\widehat{y_i}|\hat{z})
&=[p_0\sum_{k=1}^{K}{\omega_i^kN(\mu_i^{(k},\sigma_i^{2(k)})}
\\&+p_1\sum_{m=1}^{M}{\omega_i^mLap(\mu_i^{(m)},\sigma_i^{2(m)})}
\\&+p_2\sum_{n=1}^{N}{\omega_i^nLog(\mu_i^{(n)},\sigma_i^{2(n)})}\ ]
*U(-\frac{1}{2},\frac{1}{2})
\\&=c(\widehat{y_i}+\frac{1}{2})-c(\widehat{y_i}-\frac{1}{2}),
\end{split}
\end{equation}
where $N(\cdot)$ represents the uniform distribution, $Lap(\cdot)$ represents the Laplacian distribution, $Log(\cdot)$ represents the Logistic distribution, and each distribution has four related parameters \emph{i.e.} $p_i,\omega_i, \mu_i, \sigma_i$, and the value of $K$, $M$, and $N$ are set to 3. $c(\cdot)$ is the cumulative function.

\subsection{Loss Function}
As mentioned in Subsection \ref{GIC}, the commonly used loss function of GAN-based image compression scheme contains RDO loss and GAN loss. In this paper, the objective function is defined as follows:

\begin{equation}
\mathcal{L}_{(E,H,G)}={E}[d(x,\hat{x} )+\lambda(R(\hat{y} )+R(\hat{z}))+\beta logD(G(y),y)],
\end{equation}
\begin{equation}
\mathcal{L}_D={E}[-log(1-D(\hat{x},y))]+E[-log(D(x,y))],
\end{equation}
where $G$ and $D$ represent the generator and discriminator. $R(\hat{y})$ and $R(\hat{z})$ are the entropy of the latent representation and hyper latent representation. 
$d(x,\hat{x})$ denotes the distortion term, which is a combination of MS-SSIM and DISTS.  
\begin{equation}
\mathcal{L}_d^{ms}={E}[1-MS-SSIM(x,\hat{x})],
\end{equation}
\begin{equation}
\mathcal{L}_d^{dists}={E}[DISTS(x,\hat{x})],
\end{equation}
\begin{equation}
\mathcal{L}_d=k\underline{~} ms\ \mathcal{L}_{ms}+k\underline{~} di\ \mathcal{L}_{dists},
\end{equation}
\\
where $k\underline{~}{ms}$ and $k\underline{~}{di}$ are two hyperparameters, used for balancing $\mathcal{L}_{ms}$ and $\mathcal{L}_{dists}$. %The reasons for selecting MS-SSIM and DISTS are as follows:
% \\
%\textbf{MS-SSIM.} 

Different from MSE, MS-SSIM not only considers the differences between pixel levels but also incorporates structural information of images at multiple scales. This makes MS-SSIM particularly effective in evaluating the quality degradation of images, especially for structural degradation. As aforementioned, the main degradation in GAN-based reconstructed images is structure degradation. Hence, it is quite suitable for measuring the degradation that exists in GAN-based reconstructed images. 
%Due to its capacity for comprehensive image quality evaluations, MS-SSIM is well-suited for application in GAN training, which often involves the generation of large amounts of noise. By leveraging MS-SSIM, it is possible to attain more stable GAN training and generate images of higher quality.
% \\
%\textbf{DISTS.} 

DISTS is a newly developed metric for evaluating the perceptual quality of images, taking  various image characteristics, including color, texture, and structure into account. Through a detailed analysis of these features, DISTS accurately quantifies image similarity and generates more precise evaluation results. Unlike other metrics, DISTS also incorporates local texture information, which allows for a more comprehensive assessment of image quality by evaluating both global and local characteristics. Its robustness to texture and high correlation with human perception makes it suitable for measuring the degradation of reconstructed images in RDO as well.

\section{EXPERIMENTS}

\subsection{Training Setting}

\textbf{Dataset.} The COCO2017 dataset is used to train our model. It contains 118,282 images with different resolutions. At the beginning of the training, each image is scaled to the corresponding proportion and randomly cropped to the size of 256 $\times$ 256. Besides, all these data are normalized in [-1,1] before they are trained.
\\
\textbf{Training Parameters.}
In this paper, we use $\lambda$ to control the rate. We set $\lambda=\left\{2, 1, 0.5\right\}$ and obtained compressed images with three different bpp. %To maintain the stability during the training, we set $\lambda$ to the twice initial data for the first 50,000 items, and then restore it to the original value for the subsequent training items. 
We select the AMD optimizer with a batch size of 4 during training.
Besides, for the rest of hyper-prameters, we set $k\underline{~}ms$ = 765 $\times$ $2^{-5}$, $k\underline{~}{di} = 1$, and $\beta$ = 0.15
\\
\textbf{Training Strategy.}
As shown in Eq. \eqref{GAN-loss}, the loss function consists of five components. It is difficult to achieve global optimization for each component through end-to-end training. Besides, it's also challenging to achieve a good balance among different components. Here, we adopt a two-stage training strategy.

In the initial stage, we train the fundamental compression components, which include the encoder, generator, and entropy coding model. The loss function in this stage is the rate-distortion optimization function, \emph{i.e.} $\mathcal{L}_{E,H,G}=d(x,G(y))+R(\hat{y})+R(\hat{z})$. The second stage is based on the previous one, the weights of all the components above are imported for fine-tuning. In the second stage, we train the whole framework of the proposed method (including the discriminator). Meanwhile, the GAN loss is incorporated with the rate-distortion optimization function to help with training. In addition, the training process follows the rule that training discriminator and generator in an alternate way.

Such a training strategy can avoid optimizing five components at the same time. In the second stage, it is no longer prepared from scratch when training the discriminator, which means the generator has acquired a certain capability to generate relevant images.

\begin{figure*}[t]
\centering
\includegraphics[width=12cm]{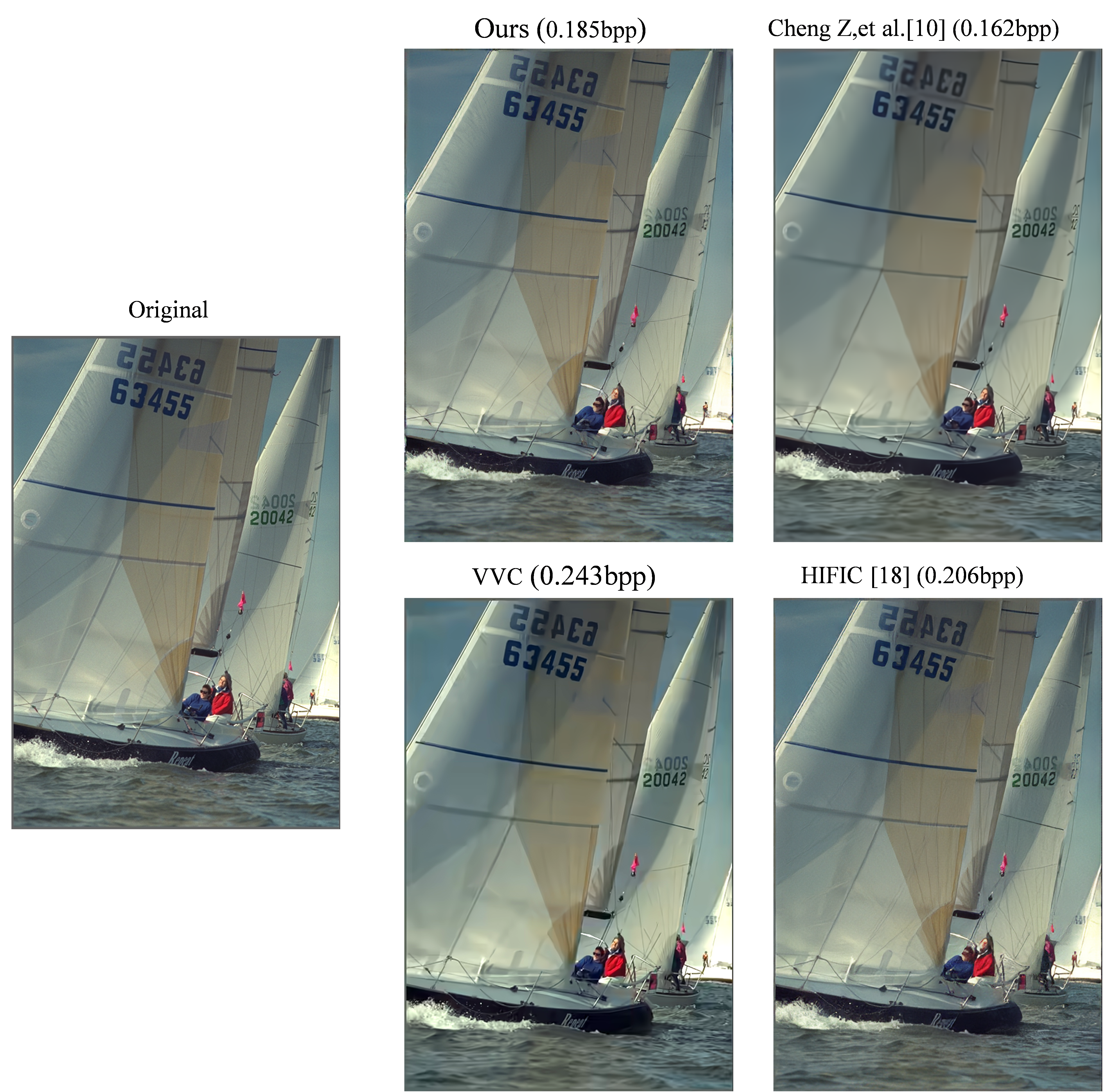}
\caption{Qualitative results of $kodim10$ from Kodak dataset}
\label{kodak10}
\end{figure*}

\subsection{Experiment Setting}
To better reflect the actual perceptual quality of users, we set a subjective test based on MOS (Mean Opinion Score) experiments. MOS experiments are a standardized method that evaluates image quality through human visual perception, thus it is believed the most convincing test for subjective quality assessment. For more details on the setting of the experimental environment, refer to the ITU-R BT.500-11 criterion \cite{bt2002methodology}.

We tested our method against the three anchors (including the state-of-the-art GAN-based image compression method \cite{mentzer2020high}, the state-of-the-art hybrid image codec VVC-4:2:0, and a commonly used learning-based image codec method \cite{cheng2020learned}) at the Kodak dataset, which contains 24 images. To this end, we compress all these 24 images with four different image codecs mentioned above under three different bpp (low, mid, and high). Then, we release 4 image pairs to each subject per test. For each image pair, the original image and its corresponding reconstructed image (generated with the image codec mentioned above) are exhibited on the right and left of the screen. All the compressed images in the 4 image pairs above have the same bpp. Then, the subject is asked to rate the score of the reconstructed image with reference to the original image. The score ranges from 1 (best) to 4 (worst). Hence, there are 24$\times$3 tests for each subject. To minimize the potential influence of other factors, we maximized the screen size of these two images. A total of 13 subjects take part in the experiment and the average score of all 13 subjects is used for comparison, namely MOS. The detailed MOS scores are exhibited in Subsection \ref{result}. 

\begin{table}[htbp]
\renewcommand\arraystretch{1.1}
  \centering
  \caption{The MOS of Kodak Dataset. ``Mean'' denotes the average of the scores, and ``Std'' is its associated standard deviation.}
  \scalebox{0.7}
  {
    \begin{tabular}{p{0.8cm}|p{1cm}|p{1.5cm}|p{1.6cm}|p{1cm}|p{1cm}|p{1.5cm}|p{1.6cm}|p{1cm}|p{1cm}|p{1.5cm}|p{1.6cm}|p{1cm}}
    \toprule
          & \multicolumn{4}{l|}{Low(0.23bpp)} & \multicolumn{4}{l|}{Mid(0.33bpp)} & \multicolumn{4}{l}{High(0.48bpp)} \\
    \hline
          & Ours   & Cheng \cite{cheng2020learned} & HIFIC \cite{mentzer2020high} & VVC   & Ours   & Cheng \cite{cheng2020learned} & HIFIC \cite{mentzer2020high} & VVC   & Ours   & Cheng \cite{cheng2020learned} & HIFIC \cite{mentzer2020high} & VVC \\
    \hline
    P1    & 1.67  & 2.88  & 2.04  & 3.42  & 2.04  & 2.79  & 2.21  & 2.96  & 1.79  & 2.46  & 2.21  & 3.54  \\
    \hline
    P2    & 2.17  & 2.50  & 2.42  & 2.92  & 1.83  & 2.88  & 2.29  & 3.00  & 1.96  & 2.71  & 2.25  & 3.08  \\
    \hline
    P3    & 2.04  & 2.58  & 2.42  & 2.88  & 2.13  & 2.88  & 2.29  & 2.71  & 1.88  & 2.33  & 2.42  & 3.38  \\
    \hline
    P4    & 2.21  & 2.58  & 2.08  & 3.13  & 1.71  & 2.33  & 2.58  & 3.38  & 1.71  & 2.08  & 2.58  & 3.63  \\
    \hline
    P5    & 2.04  & 2.75  & 2.08  & 3.13  & 1.88  & 2.63  & 2.00  & 3.50  & 1.75  & 2.42  & 2.13  & 3.71  \\
    \hline
    P6    & 1.67  & 2.88  & 1.79  & 3.58  & 1.88  & 2.71  & 1.71  & 3.67  & 1.58  & 2.67  & 2.00  & 3.67  \\
    \hline
    P7    & 2.08  & 2.42  & 2.38  & 3.13  & 1.79  & 2.58  & 2.50  & 3.13  & 1.79  & 2.63  & 2.79  & 2.79  \\
    \hline
    P8    & 2.17  & 2.54  & 2.29  & 3.00  & 2.13  & 2.71  & 2.17  & 3.00  & 2.17  & 2.46  & 2.29  & 3.08  \\
    \hline
    P9    & 1.54  & 2.17  & 2.38  & 3.92  & 1.83  & 1.96  & 2.25  & 3.96  & 1.88  & 2.83  & 2.04  & 3.25  \\
    \hline
    P10   & 1.83  & 2.63  & 2.17  & 3.38  & 1.92  & 2.58  & 2.33  & 3.21  & 1.83  & 2.25  & 2.38  & 3.54  \\
    \hline
    P11   & 1.96  & 2.88  & 2.13  & 3.04  & 1.67  & 2.79  & 2.42  & 3.13  & 1.88  & 2.83  & 2.04  & 3.25  \\
    \hline
    P12   & 1.83  & 2.71  & 2.25  & 3.21  & 1.83  & 2.50  & 2.25  & 3.42  & 1.75  & 2.25  & 2.50  & 3.50  \\
    \hline
    P13   & 1.88  & 2.75  & 2.21  & 3.17  & 1.96  & 2.54  & 2.25  & 3.25  & 1.83  & 2.50  & 2.33  & 3.33  \\
    \hline
    Mean  & 1.93  & 2.63  & 2.20  & 3.22  & 1.89  & 2.61  & 2.25  & 3.25  & 1.83  & 2.49  & 2.30  & 3.37  \\
    \hline
    Std   & 0.21  & 0.20  & 0.17  & 0.28  & 0.14  & 0.24  & 0.21  & 0.32  & 0.13  & 0.22  & 0.22  & 0.26  \\
    \bottomrule
    \end{tabular}
    }
  \label{tab2}
\end{table}

\begin{table}[htbp]
\renewcommand\arraystretch{1.1}
  \centering
  \caption{The average MOS of different methods at three different bbp.}
% \scalebox{0.51}
{
    \begin{tabular}{p{1cm}|p{1.3cm}|p{1.8cm}|p{1.8cm}|p{1.3cm}}
    \toprule
          & Ours   & Cheng \cite{cheng2020learned} & HIFIC \cite{mentzer2020high} & VVC \\
    \hline
    P1    & 1.83  & 2.71  & 2.15  & 3.31  \\
    \hline
    P2    & 1.99  & 2.69  & 2.32  & 3.00  \\
    \hline
    P3    & 2.01  & 2.60  & 2.38  & 2.99  \\
    \hline
    P4    & 1.88  & 2.33  & 2.42  & 3.38  \\
    \hline
    P5    & 1.89  & 2.60  & 2.07  & 3.44  \\
    \hline
    P6    & 1.71  & 2.75  & 1.83  & 3.64  \\
    \hline
    P7    & 1.89  & 2.54  & 2.56  & 3.01  \\
    \hline
    P8    & 2.15  & 2.57  & 2.25  & 3.03  \\
    \hline
    P9    & 1.75  & 2.32  & 2.22  & 3.71  \\
    \hline
    P10   & 1.86  & 2.49  & 2.29  & 3.38  \\
    \hline
    P11   & 1.83  & 2.83  & 2.19  & 3.14  \\
    \hline
    P12   & 1.81  & 2.49  & 2.33  & 3.38  \\
    \hline
    P13   & 1.89  & 2.60  & 2.26  & 3.25  \\
    \hline
    Mean   & 1.88  & 2.58  & 2.25  & 3.28  \\
    \hline
    Std   & 0.21  & 0.20  & 0.17  & 0.28  \\
    \bottomrule
    \end{tabular}}
  \label{tab3}%
\end{table}%

\subsection{Experimental Results}
\label{result}
\begin{figure}[t]

\centering
\includegraphics[width=12cm]{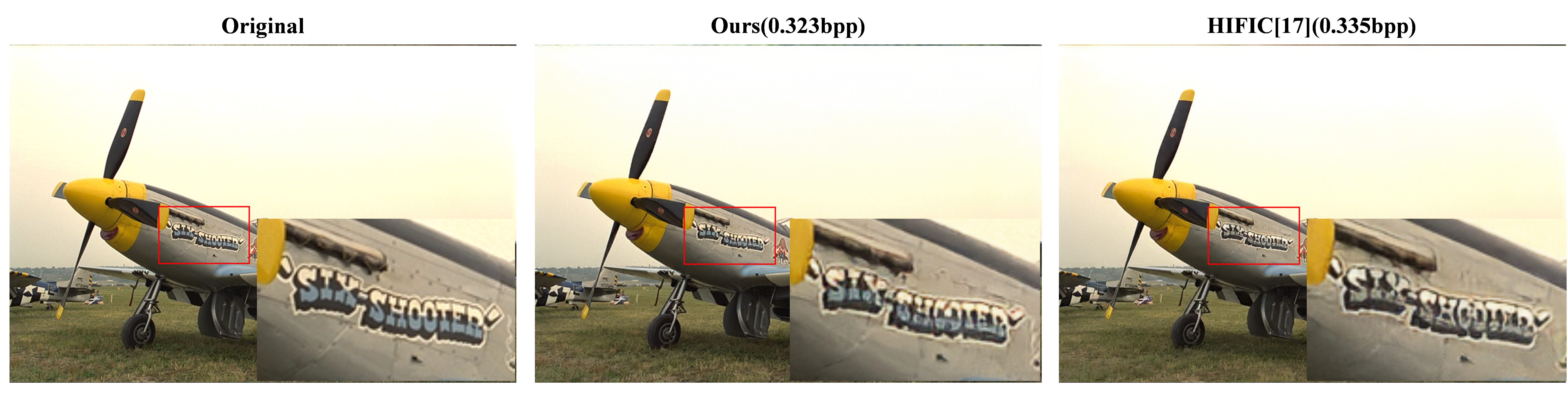}
\caption{Qualitative results of $kodim20$ from Kodak dataset}
\label{kodak20}
\end{figure}

\textbf{MOS.} The result of the MOS is shown in Table \ref{tab2} and Table \ref{tab3}. A lower score indicates a better perceptual quality of the reconstructed image. It can be seen that our method achieves the best results in the low, mid, and high bpp, respectively. %In addition, our method achieves the best results in 37 out of 39 data sets generated by the 13 participants, covering three different modes. 
The results above demonstrate the proposed method is able to achieve a better perceptual quality image reconstruction under the same bpp.
\\
\textbf{Detail Comparison.} Fig. \ref{kodak10} shows the original image of \emph{kodim10}, it associated reconstructed images with different methods and bpp. Our method outperforms the anchor methods in reconstructing detailed texture and structure, as shown in Fig. \ref{kodak10}, particularly in the region of sailboats and seawater. Additionally, Fig. \ref{kodak20} also indicates that our method achieves good performance in color fidelity compared with the state-of-the-art GAN-based method HIFIC \cite{mentzer2020high}, such as the text on the plane, which should be blue instead of gray. %Our reconstruction more accurately reproduces the original hue, while the picture reconstructed by HIFIC \cite{mentzer2020high} suffers from pronounced color distortions.

\section{CONCLUSION}
In this paper, we have optimized the state-of-the-art GAN-based image codec by improving its RDO process. To achieve this, we integrate the DISTS and MS-SSIM metrics to measure perceptual degeneration in color, texture, and structure. Meanwhile, we also absorb the GLLMM to improve the accuracy of entropy modeling, which further promotes the probability distribution estimation of the latent representation. To evaluate that the proposed method can achieve higher perceptual quality image reconstruction, the MOS experiment is performed among different codecs, which is believed as the most accurate subjective test experiment that highly reflects the actual perceptual quality of human vision. Experimental results show that our method outperforms the state-of-the-art GAN-based image codecs and hybrid image one (\emph{e.g.}, VVC).

\bibliographystyle{plain}
\bibliography{references}

\end{document}